\documentstyle[times,epsfig,graphics,amssymb]{mn2e}

\newcommand{\be}{\begin{equation}}
\newcommand{\e}{\end{equation}}
\newcommand{\bear}{\begin{eqnarray}}
\newcommand{\ear}{\end{eqnarray}}

\newcommand{\de}{{\rm d}}

\def\be{\begin{equation}}
\def\ee{\end{equation}}

\def\gsim{\lower.5ex\hbox{\gtsima}}
\def\lsim{\lower.5ex\hbox{\ltsima}}
\def\gtsima{$\; \buildrel > \over \sim \;$}
\def\ltsima{$\; \buildrel < \over \sim \;$}
\def\prosima{$\; \buildrel \propto \over \sim \;$}
\def\gsim{\lower.5ex\hbox{\gtsima}}
\def\lsim{\lower.5ex\hbox{\ltsima}}
\def\simgt{\lower.5ex\hbox{\gtsima}}
\def\simlt{\lower.5ex\hbox{\ltsima}}
\def\simpr{\lower.5ex\hbox{\prosima}}

\title[Testing Reionization with GRBs]
{Testing Reionization with Gamma Ray Burst Absorption Spectra}

\author[S. Gallerani, R. Salvaterra, A. Ferrara, T. Roy Choudhury]{
S. Gallerani$^{1}$, R. Salvaterra$^{2}$, A. Ferrara$^{1}$, T. Roy Choudhury$^{3}$ \\\\
$^1$ SISSA/International School for Advanced Studies, via Beirut
2-4, 34014 Trieste, Italy\\
$^2$ Dipartimento di Fisica G. Occhialini, Universit\`a degli Studi di Milano Bicocca, Piazza della Scienza 3, I-20126 Milano, Italy \\
$^3$ Institute of Astronomy, Madingley Road, Cambridge CB3 OHA, UK\\}

\date{\today}
\pagerange{\pageref{firstpage}--\pageref{lastpage}} \pubyear{2007}
\begin{document}

\maketitle
\label{firstpage}

\begin{abstract}
We propose to study cosmic reionization using absorption line spectra of high-redshift Gamma Ray Burst (GRB) afterglows. We show that the statistics of the dark portions (gaps) in GRB absorption spectra represent exquisite tools to discriminate among different reionization models.  We then compute the probability to find the largest gap in a given width range $[W_{\rm max}, W_{\rm max}+dW]$ at a flux threshold $F_{\rm th}$ for burst afterglows at redshifts $6.3 \le z \le 6.7$. 
We show that different reionization scenarios populate the $(W_{\rm max},F_{\rm th})$
plane in a very different way, allowing to distinguish among different reionization histories.
We provide here useful plots that allow a very simple and direct comparison 
between observations and model results. Finally, we apply our methods to GRB~050904 detected at $z=6.29$. We show that the observation of this burst strongly favors reionization models which predict a highly ionized intergalactic medium at $z\sim 6$, with an estimated mean neutral hydrogen fraction $x_{\rm HI}=6.4\pm 0.3 \times 10^{-5}$ along the line of sight towards GRB~050904.
\end{abstract}

\section{Introduction}
In the last few years, our knowledge of the high-$z$ Universe and in
particular of the reionization process has been enormously increased mainly 
owing to the observation of quasars by the SDSS survey (Fan 2006) and CMB
data (Hinshaw et al. 2007, Page et al 2007). Long gamma ray bursts (GRB) may constitute a complementary way 
to study the reionization process avoiding the proximity effects and 
possibly probing even larger redshifts. This has now become clear after the detection of five GRBs at $z\gsim 5$, 
over a sample of about 200 objects observed with the {\it Swift} satellite 
(Gehrels et al. 2004). The current
record holder is GRB~050904 at $z = 6.29$ (Tagliaferri et al. 2005, Kawai et 
al. 2006). Totani et al. (2006) have used this object to constrain the ionization state
of the intergalactic medium (IGM) at high redshift by modeling its optical 
afterglow spectrum. They report the evidence that 
the IGM was largely ionized already at $z=6.3$. The best-fit neutral 
hydrogen fraction is consistent with zero with upper limit $x_{\rm HI}<0.17$ 
($<0.6$) at 68\% (95\%) C.L.

Several authors (Natarajan et al. 2005, Daigne, Rossi \& Mochkovitch 2006, 
Bromm \& Loeb 2006, Salvaterra et al. 2007a) have computed the number of 
high--$z$ GRBs detectable by {\it Swift}. In spite of model details,
all these different studies consistently predict that a non-negligible fraction 
(up to $\sim 10$\%) of all observed GRBs should lie at very high redshift. 
On the basis of these results, several GRBs at $z\ge 6$ should be 
observed by {\it Swift} in the near future. Depending on the 
{\it Swift}/BAT trigger sensitivity and on model details, Salvaterra \& 
Chincarini (2007) found that $\sim 2-8$ GRBs can be detected above this 
redshift during every year of mission. From the observational point of view, Ruiz-Velasco et al. (2008) find that the fraction of {\it Swift} GRBs at $z>6$ should not exceed the conservative upper limit of $19\%$.
Moreover, future X-- and Gamma--ray missions will increase rapidly the sample
of high--$z$ GRBs, possibly up to $z\sim 10$ (Salvaterra et al. 2007b).
Once detected, spectroscopic follow-up observations of high--$z$ GRBs require 
a rapid trigger of  8-meter, ground based telescopes. This can be done by
pre-selecting reliable candidates on the basis of some promptly-available 
information provided by {\it Swift}, such as burst duration, photon flux, 
the lack of detection in the UVOT $V$-band, and the low Galactic extinction 
(Campana et al. 2007, Salvaterra et al. 2007a).
We note that cosmological time dilation helps keeping the flux bright, since observations will sample the afterglow early phases, even a
few days after trigger.   

The spectra of high-redshift sources (as QSOs and GRBs) bluewards of the Ly$\alpha$ are characterized 
by dark portions (gaps) produced by intervening neutral hydrogen along the line of sight. 
The use of various gap statistics in QSO spectra has been recently recognized as a very powerful 
tool to constrain the IGM ionization state (Paschos \& Norman 2005; Fan et al.
2006; Gallerani, Choudhury \& Ferrara 2006, hereafter G06; Gallerani et al. 
2007, hereafter G07). For example,  
by comparing the statistics of these spectral features in a sample of 17 observed QSOs with Ly$\alpha$ 
forest simulations, G07 concluded that the HI fraction, $x_{\rm{HI}}$, evolves smoothly 
from $10^{-4.4}$ at $z=5.3$ to $10^{-4.2}$ at $z=5.6$, with a robust upper limit $x_{\rm{HI}} < 
0.36$ at $z=6.3$. These results encourage the application of such analysis to GRBs.  
There are several advantages promised by such an attempt. First, GRBs are soon expected to
be found at redshifts higher than those typical of QSOs; second, and contrary to the massive 
hosts of QSOs, they reside in ``average" cosmic regions only marginally affected by local 
ionization effects or strong clustering; finally, they are bright and their afterglow spectra 
closely follows a power-law, making continuum determination much easier

In agreement with the WMAP3 results 
(Spergel et al. 2007) we assume a flat universe with $\Omega_m=0.24$, 
$\Omega_{\Lambda}=0.76$, $\Omega_bh^2=0.022$, $h=0.73$. The parameters 
defining the linear dark matter power spectrum are $n=0.95$, 
$\de n/\de \ln k=0$, $\sigma_{8}=0.82$. Mpc are physical unless differently stated.
\section{GRB emission properties}
We have built a synthetic GRB afterglow emission spectrum starting from
the observed spectral energy distribution and time evolution of the most
distant GRB detected up-to-now, i.e. GRB 050904 at $z=6.29$ (Tagliaferri et al. 2005,
Kawai et al. 2006). The unabsorbed afterglow spectrum of GRB 050904 can be
parametrized as $F(\nu)\propto \nu^{\rm \alpha} t^{\rm \beta}$, with $\alpha=-1.25^{+0.15}_{-0.14}$
(Haislip et al. 2006, Tagliaferri et al. 2005). The observed temporal decay 
up to 0.5 days from burst is well described by a power-law index of 
$-1.36\pm 0.06$, followed by a plateau phase with $\beta=-0.82\pm 0.15$ lasting
until 2.6 days after burst (Haislip et al. 2006). The further 
afterglow evolution can be described assuming $\beta=-2.4\pm0.4$ 
(Tagliaferri et al. 2005). Finally, we normalize the intrinsic GRB 050904 
optical spectrum in order to reproduce the flux of 
$\sim 18$ $\mu$Jy as measured at 1 day from burst in the J band (Haislip et al.
2006). Although we have experimented with other choices to explore the model
sensitivity to redshift, we will mostly show results for GRBs located in the range 
$6.3 \le z \le 6.7$, which appears to be crucial to properly
follow the latest (i.e. overlap) phases of reionization history. 
From the observed afterglow spectral evolution of GRB 050904,
we compute the rest frame spectrum between Ly$\alpha$ ($1215.67$ \AA) and Ly$\beta$ ($1025.72$ \AA). 
\section{Absorption spectra}
\begin{figure}
\centerline{
\psfig{figure=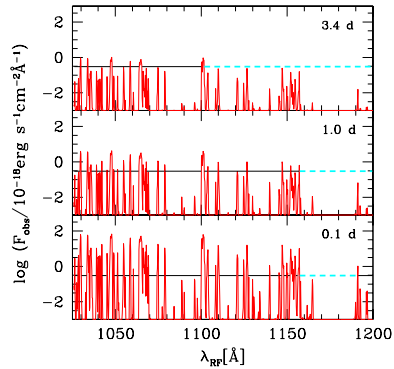,width=5.7cm,angle=0}}
\caption{Synthetic absorption spectra at different times after the burst (t=0.1, 1.0, 3.4 days from the bottom to the top) of a GRB at $z=6.3$, for the ERM. The thin black horizontal lines denote the flux threshold $F_{\rm th}^{\rm obs}$ used to define gaps. The thick dashed cyan lines show the largest dark gap along each lines of sight.}    
\label{Fig1}
\end{figure} 
The ultraviolet radiation emitted by a GRB can suffer resonant Ly$\alpha$ scattering as it propagates
through the intergalactic neutral hydrogen. In this process, photons are removed from the line of
sight (LOS)  resulting in an attenuation of the source flux, the so-called Gunn-Peterson (GP) 
effect.  Thus, the observed flux $F_{\rm obs}$ is given by $F_{\rm obs}=F(\nu)~e^{-\tau_{\rm GP}}$, where $\tau_{\rm GP}$ is the GP optical depth. To simulate the $\tau_{\rm GP}$ distribution 
we use the method described by G06 and further revised in G07, whose main features are summarized as follows. 
Mildly non-linear density fluctuations giving raise to spectral absorption features in the 
intergalactic medium (IGM) are described by a Log-Normal distribution, which has been shown to fit observational data at redshifts $1.7<z<5.8$ (Becker et al. 2007).   
For a given IGM equation of state (i.e. temperature-density relation), the mean HI fraction, $x_{\rm HI}$, 
can be computed from photoionization equilibrium (an excellent approximation under the prevailing physical conditions) 
as a function of the baryonic overdensity, $\Delta\equiv\rho/\bar\rho$, and photoionization rate, $\Gamma$, due to the 
ultraviolet background radiation field. These quantities must be determined from a combination of theory and
observations; here we follow the approach of Choudhury \& Ferrara (2006, CF06 hereafter). Their study allows for three 
types of ultraviolet ionizing photons: QSOs, PopII and PopIII stars. As the contribution of PopIII stars is negligible
for the redshift of interest here, we neglect this type of sources. The CF06 model contains two free parameters: 
(i) the star-formation efficiency $f_*$, and (ii) the escape fraction $f_{\rm esc}$ of ionizing photons from galaxies. 
These are calibrated by a maximum-likelihood procedure to a broad data set, including the redshift evolution of
Lyman-limit systems, Ly$\alpha$ and Ly$\beta$ GP optical depths, electron scattering optical depth, cosmic star formation 
history, and number counts of high redshift sources.

Currently, the available data can be explained by two different reionization histories, corresponding to different choices of the free parameters: 
(i) an Early Reionization Model (ERM) ($f_{*}=0.1; f_{\rm esc}=0.07$), and (ii) a Late Reionization Model (LRM) ($f_{*}=0.08; f_{\rm esc}=0.04$).
These two models bracket the range of possible reionization histories. For example models in which reionization occurs earlier (later) than
in the ERM (LRM) would overproduce the electron scattering (GP) optical depth. 
The global properties of the two reionization models are shown in Fig. 1 of G07
so we do not report them here. It suffices to say that in the ERM the volume filling factor of ionized regions,
$Q_{\rm HII}=V_{\rm HII}/V_{\rm tot}=1$ at $z\leq 7$; in the LRM $Q_{\rm HII}$ evolves from 0.65 to unity in the redshift range 7.0-6.0, implying that the 
reionization is still in the pre-overlap stage at $z\geq 6$. Both ERM and LRM by construction provide an excellent fit to the
mean neutral hydrogen fraction evolution experimentally deduced from the GP test.
Having selected all the necessary physical 
evolutionary properties of the most likely reionization histories, we can now describe how we plan to use GRBs to discriminate
between them.  
\section{Reionization tests}
The main idea we propose in this paper is to exploit the statistics of the 
transmissivity gaps imprinted by the intervening IGM neutral hydrogen on the 
otherwise smooth power-law spectrum of high-redshift GRBs. On general grounds, 
we expect that at any given redshift, but particularly above $z=6$, where 
differences become more marked, the value of $x_{\rm{HI}}$ is higher in the 
LRM than in the ERM. As a result, if reionization completes later, absorption spectra are characterized by wider and more numerous gaps, defined as contiguous regions of the spectrum with observed 
flux lower than a given flux threshold over a rest frame 
wavelength interval larger than 1 \AA. Moreover, the level of neutral hydrogen 
fluctuations along the LOS to the GRB differs 
in the two models: in principle, the GRB flux decay could be used as a tunable 
high-pass filter which allows to study the growth
of gaps with time. 
As the time after the burst increases, the gaps become larger.
In fact, the progressive fading of the unabsorbed afterglow produces a
corresponding attenuation of the observed flux, as can be seen from Fig. 1.
As the flux falls below the flux threshold $F_{\rm th}^{\rm obs}$ (thin black horizontal lines) the
transmission
windows disappear, causing the gaps to expand. The thick dashed cyan line shows the
largest gap ($W_{\rm max}$) along the LOS; the largest gap width (LGW)
increases as the GRB flux fades.
However, in practice, the analysis of the LGW time evolution can be equivalently applied at a fixed time after the burst to the transmitted flux $F_{\rm transm}=F_{\rm obs}/F(\nu)$ varying the threshold\footnote{Note that we define $F_{\rm th}^{\rm obs}$ as the threshold which defines gaps in the observed flux and $F_{\rm th}$ in reference to the transmitted one.} $F_{\rm th}$. 

We have derived the evolution 
of the LGW found in synthetic absorption spectra varying the flux threshold used to define gaps.  
Suppose that a GRB is observed at a given redshift $z_{\rm GRB}$.
We can then ask what is the probability that in its afterglow spectrum the largest gap, defined by $F_{\rm th}$, is found within a given width range\footnote{In this work we assume $dW=20$\AA.} $[W_{\rm max}, W_{\rm max}+dW]$. 
We can draw this probability from our model by analyzing a large number of 
synthetic spectra corresponding to different GRB redshifts $z_{\rm GRB}$ and 
counting the number of LOS having the largest gap in the given range, varying the $F_{\rm th}$  between 0 and 1. Clearly, in real cases, a minimum threshold $F_{\rm th, min}$ is set by the noise associated to the spectrum.

We have computed this statistics for afterglow spectra of GRBs with $6.3 < z_{\rm GRB} < 6.7$.
In fact, according to the most
recent estimates involving the lower trigger sensitivity, the {\it Swift} mission is expected to increase the sample
of high--$z$ GRBs, possibly up to $z\sim 10$ (Salvaterra et al. 2007b). 
\begin{figure*}
\centerline{\psfig{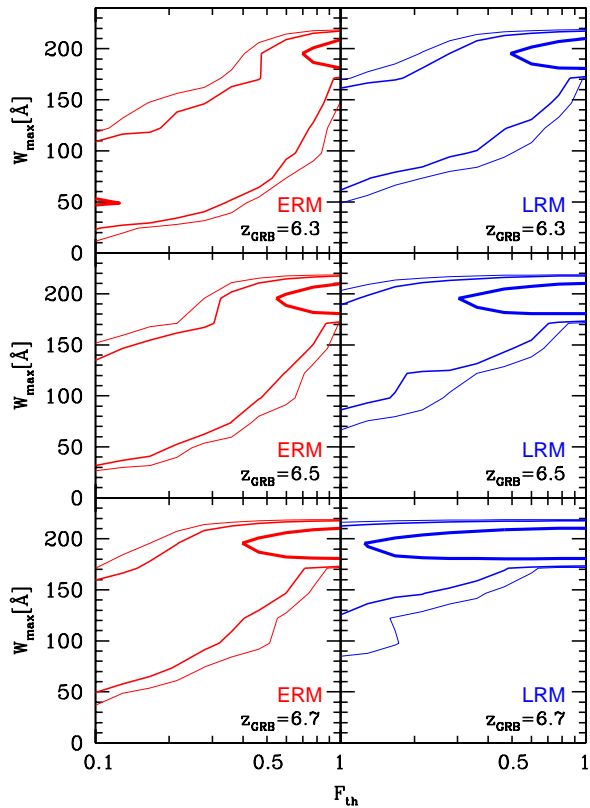}}
\caption{Isocontours of the probability that the afterglow spectrum of a GRB at redshift $z_{\rm GRB}$, contains a 
largest 
gap of size in the range $[W_{\rm max}, W_{\rm max}+dW]$, with $dW=20$~\AA for a flux threshold $F_{\rm th}$.  The left (right) panel shows the results for the ERM (LRM). The top (middle, bottom) panel shows the results for $z_{\rm GRB}=6.3$ ($6.5, 6.7$).
The isocontours correspond to probability of 5\%, 10\%, and 40\%.} 
\label{Fig1}
\end{figure*} 

We show the resulting largest gap probability isocontours in Fig. 2 for 
$z_{\rm GRB}=6.3$ (top panels),  $z_{\rm GRB}=6.5$ (middle panels), and  $z_{\rm GRB}=6.7$ (bottom panels)
; the left (right) panels refer to the ERM (LRM) case. We use 300 LOS   in order to ensure statistical convergence of the results. The isocontours correspond to a probability of 5\%, 10\%, and 40\%.  
The differences caused by the two different reionization histories are striking, since the ERM and LRM populate the  $(W_{\rm max},F_{\rm th})$
plane in a very different way, as shown in Fig. 2. For example, at $z_{\rm GRB}=6.3$, the probability to find the largest gap with $W_{\rm max}=50$~\AA~for $F_{\rm th}=0.1$, is $\approx 40\%$, in the ERM, while it is only of the $\approx 5\%$ in the LRM, being most of the largest gaps wider than 50~\AA.

Moreover, in the LRM the width of the gaps is a factor $\approx 2$  times wider 
than in the ERM.  In particular, for $z_{\rm GRB}=6.5$ and $F_{\rm th}=0.2$, we find that in the ERM $40\lesssim W_{\rm max}\lesssim 170$~\AA, while in the LRM gaps are mostly larger than $80$ \AA, for the same value of the flux threshold.  

Finally, in the ERM the LOS becomes completely dark for flux threshold values higher than in the LRM case, as a consequence of the higher transmitted flux predicted by the former. For example, at $z=6.7$, in the ERM (LRM) the gap size is equal to the simulated LOS length ($\approx 190$\AA) for $F_{\rm th}\approx 0.4~(0.15)$.

The results point also towards two very important, and potentially interesting, redshift-dependent features, both
related to the thickening of the Ly$\alpha$ forest, which becomes on average more neutral towards higher redshifts.   
First, there is an overall shift towards larger $W_{\rm max}$ values of both ERM and LRM curves from $z_{\rm GRB}=6.3$ to $z_{\rm GRB}=6.7$. Second, the
rate at which the gap width grows and saturates with $F_{\rm th}$ strongly correlates with redshift.   

Thus, the analysis of the largest gap evolution with the flux threshold shows that it is is easier to discriminate among different reionization 
histories using the highest redshift available sample of GRBs. 

Fig.~2 allows a straightforward comparison between data and model results. In practice, information 
on the reionization history from a $z>6$ GRB can be easily extracted through the following the steps:
(i) from the J-band afterglow flux, $F_{\rm J}$, extrapolate the continuum as $F(\nu)=F_{\rm J} 
(\nu/\nu_{\rm J})^{\alpha}$; (ii) compute the transmitted flux $F_{\rm transm}=F_{\rm obs}/F(\nu)$; (iii)
Determine the LGW in $F_{\rm transm}$, varying $F_{\rm th}$ in $[F_{\rm th, min};1]$, (iv) compare the result
with Fig. 2. This simple procedure will allow to dicriminate between different reionization scenarios.
\section{The case of GRB 050904} 
We apply the LGW analysis to the observed flux of the GRB 050904, whose  
spectrum has been obtained 3.4 days after the burst (Kawai et al. 2006). For a fair comparison with observations, simulated spectra have been convolved with a gaussian of
FWHM = 300 km/sec, providing $R=1000$, and then rebinned to a resolution of
R = 600. Finally, we add noise to the simulated data such that the
flux F in each pixel is replaced by F + G(1) $\sigma_n$, where G(1) is
a Gaussian random deviate with zero mean and unit variance, and $\sigma_n=0.05$ is almost the mean noise r.m.s deviation, observed in the GRB 050904 spectrum by 
Kawai et al. 2006.

In this work the lower flux threshold value that we use is
$F_{\rm th, min}^{\rm obs}=0.3\times 10^{-18}{\rm erg~s^{-1} cm^{-2} \AA^{-1}}$ which is $\geq 3-\sigma_n$. Such conservative choice of $F_{\rm th}^{\rm obs}$ excludes the 
possibility that the widths of the gaps are underestimated because of some 
spurious transmission peaks.
For $F_{\rm th}^{\rm obs}=(0.30;~0.35;~0.40)\times 10^{-18}\rm erg/\rm s/cm^2/\rm{\AA}$, 
the observed largest dark gaps sizes are $W_{\rm max}=(51;~78;~141)$~\AA, 
respectively, in the GRB rest frame.  These values refer to the dark regions 
extending in the observed wavelength region $\sim [7850-8220]$; 
$[7850-8420]$; $[7850-8874]$~\AA, respectively. It is worth noting that the 
third wavelength range contains the spectral region immediately blueward the Ly$\alpha$ 
emission, which could be affected by the presence of a DLA, as claimed by 
Kawai et al. (2006). Thus, the corresponding quoted size for the dark gap 
should be consider as an upper limit. 

The black filled circles in Fig. 3, 
corresponding to the values of $W_{\rm max}$ and $F_{\rm th}^{\rm obs}$ for GRB 050904, show that the ERM is favored by the data.  
In particular, the probability to find a largest gap of 
$W_{\rm max}\approx 51$~\AA, for 
$F_{\rm th}^{\rm obs}=0.30\times 10^{-18}\rm erg/\rm s/cm^2/\rm{\AA}$ is 
$\approx 40$\% if reionization occurred early, i.e. almost half of the LOS 
contains a largest gap of this size for a burst with the redshift and flux of 
GRB~050904. Such probability drops for a late reionization model to $\sim 10$\% 
clearly indicating that in this scenario the GRB~050904 observation represents 
a much rarer event. 
Moreover, in the ERM, the probability to find a largest gap of 
$W_{\rm max}\approx 78$~\AA, for 
$F_{\rm th}^{\rm obs}=0.35\times 10^{-18}\rm erg/\rm s/cm^2/\rm{\AA}$ is almost 3 times the LRM one (27\% vs 10\%).
Finally, for the third observed point the LRM appears to be more probable
than the ERM by a factor 27:5. This value should be considered an upper
limit, since the third point has been obtained analyzing a spectral region which could be affected
by a DLA contamination (Kawai et al. 2006). For this reason, the corresponding gap size 
should be smaller than $141$\AA, thus being more consistent with ERM 
predictions, as the other observational data. 
Without taking into account the latter 
observed point, we find that the ERM is 11 times more probable than 
the LRM; if such data point is included, the above vantage factor 
reduces to 2. We have checked the impact of uncertainties on the continuum flux
by varying its slope within the quoted error on $\alpha$ (see Sec. 2).
We have found that the results shown in Fig. 3 are negligibly affected.\\
Our results are due to the fact that many, relatively large, opaque 
stretches are present in the IGM at $z=6.3$ for the LRM, 
resulting in an average large $W_{\rm max}$. In the case of ERM, reionization 
is almost complete already at $z\sim 7$ and smaller largest gaps 
are expected at the redshift of GRB~050904. \\
Although a large sample of high redshift GRBs is required before we conclude 
that a model in which reionization ended by  $z\approx 7$ is favored by the 
data, the discriminating power of the proposed method is already apparent.
\subsection{Neutral hydrogen measurements}
We have compared the gap statistics as predicted by two different reionization models. At z=6, the ERM predicts a mean neutral hydrogen fraction $x_{\rm HI}\sim 10^{-4}$, while in the LRM $x_{\rm HI}\sim 0.03$. As shown in the previous Sections, the gap statistics is sensitive to the neutral hydrogen amount along the LOS to the background source. Thus, we can exploit the better agreement of the ERM predictions with observed data to provide a measurement of the neutral hydrogen fraction $x_{\rm HI}$ along the LOS to the GRB 050904. To this aim, from the ERM sample used in the analysis, we select those LOS whose largest gap is consistent with observations. Thus, we derive the mean neutral hydrogen fraction $x_{\rm HI}$ along the synthetic LOS characterized by $W_{\rm max}~\epsilon~[51~$\AA$~\pm~dW]$ for $F_{\rm th}^{\rm obs}=0.30\times 10^{-18}$~erg~s$^{-1}$~cm$^{-2}$~\AA$^{-1}$ and $W_{\rm max}~\epsilon~[78~$\AA$~\pm~dW]$ for $F_{\rm th}^{\rm obs}=0.35\times 10^{-18}$~erg~s$^{-1}$~cm$^{-2}$~\AA$^{-1}$. We find that the observed LGW 
in the GRB~050904 afterglow spectrum are consistent with $x_{\rm HI}=6.4\pm 0.3\times 10^{-5}$. 
\begin{figure*}
\centerline{\psfig{figure=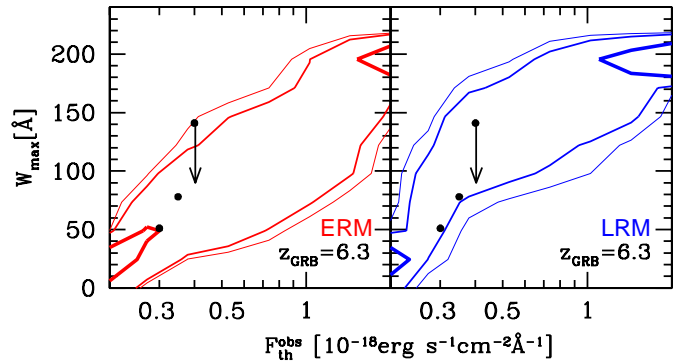,width=8.8cm,angle=0}}
\caption{Isocontours of the probability that the afterglow spectrum 
associated with a GRB at redshift $z_{\rm GRB}=6.3$, contains a 
largest 
gap of size in the range $[W_{\rm max}, W_{\rm max}+dW]$, with $dW=20$~\AA, for a flux threshold $F_{\rm th}$.  
The left (right) panel shows the results for the ERM (LRM). The isocontours correspond to probability of 5\%, 10\%, and 40\%. The black points indicates the position in the $(W_{\rm max}, F_{\rm th}^{\rm obs})$ plane of 
GRB 050904. The point with arrow means that the gap size should be considered as an upper limit, since the corresponding dark region could be affected by the presence of a DLA (Kawai et al. 2006).}
\label{Fig1}
\end{figure*} 
\section{Conclusions}
We have proposed to investigate cosmic reionization using absorption line spectra of high--$z$ GRB afterglows. 
The evolution of the largest gap width (LGW) as a function of the flux threshold  $F_{\rm th}$ used to define 
gaps shows marked differences in the two reionization models favored by present data. 
In particular, (i) the LGW is typically $\sim 2$ times wider in the LRM than in the ERM, for a fixed $F_{\rm th}$; 
(ii) the difference between ERM and LRM in terms of $(W_{\rm max}, F_{\rm th})$ increases with $z$; (iii) an overall shift towards larger $W_{\rm max}$ values for a fixed $F_{\rm th}$ is found in both models towards higher redshifts. 
This analysis has shown that we can robustly distinguish among different reionization histories: improved results can 
be obtained if data is collected promptly after burst detection and by using GRBs at the highest available redshifts.

A direct comparison of the model with data can be carried on in terms of the probability to find the largest gap in a given width range for burst afterglows at $z=z_{\rm GRB}$.  When applied 
to the only known GRB at $z>6$, i.e. GRB~050904 at $z_{\rm GRB}=6.29$, a clear indication is obtained that reionization must 
have occurred well before $z=6$. 

We find that the observed LGW 
in the GRB~050904 afterglow spectrum are consistent with $x_{\rm HI}=6.4\pm 0.3\times 10^{-5}$. This result 
is in agreement with previous measurements by Totani et al. (2006), who find that $x_{\rm HI}$ is bound to be  
$x_{\rm HI}<0.17$ ($<0.6$) at 68\% (95\%) C.L.

Some gaps/peaks in absorption spectra could be due to DLAs/HII regions intervening along the LOS towards 
the background source (e.g. the DLAs/transverse proximity effect detection by Totani et al. 2006/Gallerani et al. 2007). 
Such contaminants could affect neutral hydrogen measurements, if the foreground sources are not properly removed. 

In this work we have applied the gap statistics to the GRB 050904 observed flux. However, the same analysis 
can be done by using the transmitted flux $F_{\rm obs}/F(\nu)$, i.e. normalizing the observed flux to the continuum. Following this approach, we provide useful plots (Fig. 2) 
which allow a straightforward comparison between our model results and future observations of $z>6$ GRBs afterglow 
spectra. 

It is worth noting that the gap statistics with a variable flux threshold can be readily applied to other background sources, as QSOs. 
We plan to apply this analysis to the Fan et al. (2006) data, thus improving the results obtained by Gallerani et al. (2007). 
The advantage of varying the flux threshold is that this technique allows to study the response of the transmitted flux both to 
the high and low end tail of the IGM density field. In fact, high (low) $F_{\rm th}$ values probe regions corresponding to 
low (high) overdensities.

In spite of the fact that QSO data are presently more abundant, GRBs are expected to be found at higher redshift, as they
should result from the death of (early) massive stars (e.g. Abel et al. 2002, Schneider et al. 2002). 
Larger samples of GRB afterglow spectra at $z\ge 6$, likely available in the near future, will allow a statistically 
significant analysis and possibly to reconstruct the cosmic reionization history. At redshifts approaching reionization epoch, 
gaps in afterglow absorption spectra are expected to become as large as the spectral region between the Ly$\alpha$ and 
Ly$\beta$ emission lines, thus obscuring completely the GRBs optical counterpart. 
\section*{Acknowledgments}
We thank E. Pian for providing useful instrumental data, G. Chincarini, D. Malesani, G. Tagliaferri for
stimulating discussions about GRB afterglows and {\it Swift} data, Z. Haiman for enlightening comments, and
the referee J. Schaye for constructive criticism.

\end{document}